\documentclass[conference]{IEEEtran}
\IEEEoverridecommandlockouts
\usepackage{cite}
\usepackage{amsmath,amssymb,amsfonts}
\usepackage{algorithmic}
\usepackage{graphicx}
\usepackage{textcomp}
\usepackage{xcolor}
\usepackage[utf8]{inputenc}
\usepackage[T1]{fontenc}
\usepackage[margin=1in]{geometry}
\usepackage{amsmath,amssymb,amsthm}
\usepackage{booktabs}
\usepackage{multirow}
\usepackage{graphicx}
\usepackage{xcolor}
\usepackage{listings}
\usepackage{pgfplots}
\pgfplotsset{compat=1.17}
\usepackage{enumitem}
\usepackage[numbers,sort&compress]{natbib}
\usepackage{hyperref}
\usepackage{cleveref}

\newcommand{\bpd}{BPD}

\newtheorem{hypothesis}{Hypothesis}

\crefname{hypothesis}{}{}

\lstdefinestyle{trace}{
  language=Python,
  basicstyle=\ttfamily\small,
  commentstyle=\color{teal!70!black}\itshape,
  keywordstyle=\color{blue!60!black}\bfseries,
  showstringspaces=false,
  columns=fixed,
  keepspaces=true,
  basewidth=0.52em,
  frame=single,
  xleftmargin=2em,
  numbers=left,
  numberstyle=\tiny\color{gray}
}
\def\BibTeX{{\rm B\kern-.05em{\sc i\kern-.025em b}\kern-.08em
    T\kern-.1667em\lower.7ex\hbox{E}\kern-.125emX}}
\begin{document}

\title{On the Limits of Execution Traces for Repairing Assertion Failures in LLM-Generated Code
\thanks{This work is funded under the agreement with the ACT Government, Future Jobs Fund - Open Source Institute(OpenSI) - R01553; and NetApp Technology Alliance Agreement with OpenSI - R01657.
Additionally, the Australian Government supported this research through the Department of Education’s National Industry PhD Program (project 36337). 
The views expressed herein are those of the authors and are not necessarily those of the Australian Government or the Department of Education.}
}

\author{\IEEEauthorblockN{Muntasir Adnan\IEEEauthorrefmark{1}}
\IEEEauthorblockA{\textit{Open Source Institute} \\
\textit{Faculty of Science and Technology}\\
\textit{University of Canberra}\\
Canberra, Australia \\
\thanks{\IEEEauthorrefmark{1}Corresponding author: Adnan.adnan@canberra.edu.au}
~\\
\and
\IEEEauthorblockN{Carlos C. N. Kuhn}
\IEEEauthorblockA{\textit{Open Source Institute} \\
\textit{Faculty of Science and Technology}\\
\textit{University of Canberra}\\
Canberra, Australia}
}
}

\maketitle

\begin{abstract}
Iterative repair returns a failed program's error to a large language model (LLM) and asks it to fix it.
After iterative repair, the remaining failures are mostly assertion failures: with four Qwen2.5 models, they make up 79\% of residual failures on BigCodeBench, all on HumanEval and 93\% on LiveCodeBench.
We argue, through an information view of the feedback, that an assertion failure says little about where the fault lies.
We propose Breakpoint Debugging (\bpd{}), which returns the failed program annotated inline with its variables' runtime values.
We take the tasks that iterative repair failed to fix with five repair attempts and compare two ways of spending five further generations on them: \bpd{}, and resampling.
On BigCodeBench, \bpd{} repairs 26.7\% of these tasks against 22.4\%, and its advantage grows with model size.
On HumanEval, \bpd{} outperforms resampling but only slightly, and on LiveCodeBench, fresh sampling repairs more (6.5\% against 4.9\%).
Furthermore, \bpd{} reduces unchanged-code resubmissions from 28\% to 8\% on BigCodeBench. 
Using only the first failing test's trace, it outperforms tracing all failing tests while using fewer tokens than both all-trace \bpd{} and plain iterative repair.
On LiveCodeBench, \bpd{} gives no gain even on the public tests, so a visible failing input is not enough for a trace to help.
These results match the information view: a trace helps only when the test exposes a failing input, and the fault is an implementation slip; otherwise, a repair loop should resample rather than trace.
\end{abstract}


\section{Introduction}
\label{sec:intro}
 
Programs written by large language models (LLMs) often fail their tests, and on benchmarks such as BigCodeBench and LiveCodeBench, even strong models solve only part of the tasks on the first attempt~\cite{chen2021codex,zhuo2025bigcodebench,jain2025livecodebench}.
Iterative self-debugging executes the program, returns the error to the model and asks for a corrected version~\cite{chen2024selfdebug, adnan2025pycapsule, zhong2024ldb}, and the quality of this feedback is widely seen as its bottleneck~\cite{olausson2024selfrepair}.

Previous studies take opposite positions on how rich the feedback should be.
LDB~\cite{zhong2024ldb} segments the program into basic blocks, records intermediate variable values and asks the model to verify each block for every type of failure.
PyCapsule~\cite{adnan2025pycapsule}, MapCoder~\cite{mapcoder} return a lightly processed error message that keeps the relevant traceback frames, including expected and actual values for assertion failures.
With the same LLM, PyCapsule reported 85.2\% against LDB's 82.9\% on HumanEval and 78.5\% against 76.0\% on MBPP~\cite{mbpp}.
These figures are not a controlled comparison, but recent studies point the same way: execution traces in repair prompts beat trace-free prompts in only two of six configurations in one study~\cite{haque2025traces} and were of limited use in another~\cite{wang2025semantics}, while trace-driven agents report gains~\cite{huang2026tracecoder, wu2026tracerepair}.
What is missing is an account of when a trace should help and when it cannot.

We start from the failures that a structured-error loop leaves behind, and find that they are almost all assertion failures.
Unlike crashes, which often indicate where execution first goes wrong, assertion failures identify only where an incorrect result is detected, offering little information about the underlying fault~\cite{voas1992pie}. 
We formalise this information gap in bits and derive five hypotheses (Section~\ref{sec: theory}).
We then propose \bpd{}, which targets assertion failures and returns the program with the values from the failing run written as inline comments, similar to the inline traces of NExT~\cite{ni2024next} but used at inference time without training.
 
This paper makes three contributions.
We show that assertion failures make up 79--100\% of the failures left by iterative repair, and explain with an information view why they resist repair.
We propose \bpd{}, which adds inline state traces only to assertion failures, and compare it with fresh sampling across four model sizes, measuring repairs, repair behaviour, and token cost.
We identify, from LiveCodeBench, when traces cannot help, and turn this into practical rules for when a repair loop should trace or resample.

We organise the evaluation around five research questions.
\textbf{RQ1:} Which failures remain after iterative repair?
\textbf{RQ2:} Do traces repair assertion failures better than fresh sampling?
\textbf{RQ3:} How do traces change repair behaviour?
\textbf{RQ4:} Is one trace enough, and what do traces cost?

\section{Related Work}
\label{sec: related}

\textbf{Iterative self-repair.}
Self-Debugging~\cite{chen2024selfdebug} showed that LLMs can improve their code from execution results and their own explanations, and PyCapsule~\cite{adnan2025pycapsule} made the loop cheaper with structured error messages.
Without external feedback, LLMs struggle to correct their own reasoning~\cite{huang2024cannot}, and once you account for repair costs, self-repair often yields modest gains limited by feedback quality~\cite {olausson2024selfrepair}.
 
\textbf{Execution information as feedback.}
Several methods give the model runtime information instead of an error message alone.
LDB~\cite{zhong2024ldb} splits a program into basic blocks and asks the model to verify each intermediate state, and NExT~\cite{ni2024next} trains models to reason over traces written as inline comments, the format closest to \bpd{}.
Recent agentic systems collect traces and pass them to several cooperating or debating agents~\cite{huang2026tracecoder,wu2026tracerepair,dynafix2025}.
However, empirical evidence for their benefits is mixed: traces outperform trace-free baselines in only two of six prompt configurations, with smaller gains from more complex traces~\cite{haque2025traces}, while another study finds limited utility for trace-based information in both fine-tuning and inference~\cite{wang2025semantics}.
LLMs are also unreliable at predicting program execution themselves~\cite{gu2024cruxeval}, which is the gap a trace is meant to fill.
 
\textbf{Repair versus resampling.}
Repair spends its budget refining one program, whereas resampling spends it on new ones.
Tang et al.~\cite{tang2024rex} frame this as an exploration--exploitation trade-off, and Zheng et al.~\cite{zheng2025multiturn} find that multi-turn repair can underperform repeated sampling on competitive programming, where problem understanding rather than repair limits performance.
 
Our work differs in three respects.
We apply traces only to assertion failures, where the information view predicts they add the most; we compare trace-guided repair with fresh sampling at an equal budget; and we explain, rather than only report, when traces cannot help.

\section{Assertion Failure: An Information View}
\label{sec: theory}


 
Let $P$ be a candidate program, $x$ a failing test input and $y^\ast$ its expected output.
Running $P$ on $x$ produces a trace $\tau = (s_0, \dots, s_T)$, where $s_t$ is a program state consisting of the values of all live variables at step $t$.
The run either raises an exception at some step or returns $y = s_T$, which the test compares with $y^\ast$.
We model the fault as a single unknown line $L$ with a prior spread over $n$ lines.
Its entropy is bounded by $H(L) \leq \log_2 n$ bits, with equality under a uniform prior.
The model receives feedback $o$, and we measure how informative $o$ is about the fault by the mutual information
\begin{equation*}
    I(L; o) = H(L) - H(L \mid o)
\end{equation*}
the number of bits by which $o$ reduces the uncertainty about $L$.
Under a uniform prior, $n$ candidate lines contain $\log_2 n$ bits of uncertainty. Thus, 5 bits are sufficient in principle to distinguish among 32 equally likely candidates. An observation carrying one bit of mutual information reduces the posterior entropy by one bit on average.

\subsection{Four differences between crashes and assertion failures}
 
\paragraph{Who judges the failure}
A crash violates a rule that the interpreter already knows, such as an index out of range, so the interpreter acts as a judge that needs no knowledge of the programmer's intent.
An assertion failure violates no rule of the language.
A function that tests for odd numbers when the task asked for even numbers runs without error and returns the wrong total, and only the expected value of the test reveals the error.
 
\paragraph{How far the symptom is from the cause}
A fault produces an observable failure only if three things happen in turn: the faulty line is executed, it corrupts the program state (infection), and the corrupted state reaches the output (propagation), as described by the PIE model~\citep{voas1992pie}.
A crash is usually reported at the moment of infection, close to the fault.
An assertion failure appears only after propagation, and its traceback names the assertion in the test, which is the same line wherever the fault is.
 
\paragraph{Information is lost, never created, along the chain}
For a fixed program and test $(x, y^\ast)$, the fault location determines the execution trace, which determines the returned value and, consequently, the verdict.
The feedback variants that a repair loop can return are nested functions of one another.
Let $b = \mathbb{I}[y = y^\ast]$ the pass/fail verdict, $y = s_T$ the returned value and $\tau$ the full trace, all for the same fixed program and failing test.
Since $L \rightarrow \tau \rightarrow y \rightarrow b$ forms a Markov chain,
\begin{equation*}
I(L; b) \;\le\; I(L; y) \;\le\; I(L; \tau),
\end{equation*}
The verdict $b$ is binary and has at most 1 bit of entropy.
Since mutual information cannot exceed the entropy of the observed variable, 
\begin{equation*}
    I(L; b) \le H(b) \le 1
\end{equation*}
The proposition follows directly from the data-processing inequality~\citep{cover2006elements}.
The three terms correspond to three feedback regimes in our experiments: a hidden test that only reports failure ($b$), a visible test whose message shows the expected and actual output ($y$), and ($\tau$).
The middle inequality is often strict because the map from traces to final values is many-to-one.
The proposition bounds what a trace could add; it does not say that the model can use it.
The information that a computationally bounded observer can actually extract, the usable or $\mathcal{V}$-information~\citep{xu2020usable}, may be much smaller than $I(L; \tau)$, and unlike Shannon information~\cite{shannon} it can decrease when more data are added, for instance when a long trace distracts the model from the relevant part~\citep{shi2023distracted,liu2024lost}.
 

\paragraph{Candidate-wise pass/fail feedback is information-inefficient} 
Suppose a repair procedure considers $N$ mutually exclusive candidate edits, with a uniform prior over the correct edit $L$. 
Identifying the correct edit requires $H(L)=\log_2 N$ bits; for $N=1000$, this is approximately $10$ bits. 
A binary verdict can convey at most one bit in a single attempt, but this upper bound is attained only when its outcomes are approximately equiprobable. 
If an attempt tests one particular candidate $e$, then \[ B=\mathbb{I}[L=e], \] so, under the uniform prior, \[ I(L;B)=H(B)=h_2(1/N), \] where $h_2$ is the binary entropy function. 
For $N=1000$, this is only approximately $0.011$ bits before observing the outcome. 
Moreover, if the attempt fails, the remaining uncertainty decreases only from $\log_2 N$ to $\log_2(N-1)$, a reduction of approximately $0.0014$ bits for $N=1000$. 
Thus, verdict-only feedback is not intrinsically uninformative: a sequence of verdicts can eventually identify the correct repair.
Rather, candidate-wise testing is information-inefficient because each unsuccessful attempt produces a highly unbalanced partition of the remaining hypothesis space. 
Effective search must therefore obtain much of its guidance from the model's prior, from the structure of the attempted edit, or from richer execution feedback.
This illustrates the limitation of verdict-only feedback, consistent with the diminishing contribution of later attempts observed in previous studies~\citep{adnan2025pycapsule, ddi} and with findings that multi-turn repair can underperform repeated sampling on competitive programming~\citep{zheng2025multiturn}.

\subsection{Why values might help a model}
 
Without a trace, a model that wants to locate a fault must simulate the program internally.
If it simulates each step correctly with probability $a$, it simulates $T$ steps correctly; with $a = 0.98$ and $T = 50$, it has only about a 36\% chance of simulating all 50 steps correctly under this simplified model.
A trace replaces simulation with lookup, so the model only has to judge whether each recorded value is plausible given the task.
This resembles the contrast between process-level and outcome-level supervision in reasoning, where feedback on intermediate steps is more useful than feedback on the final answer alone~\citep{uesato2022process,lightman2024verify}.
We derive five hypotheses from the argument above.
\begin{hypothesis}[Localisation]
\label{h: local}
Where the feedback already shows the failing input, the expected output and the actual output, a trace still adds information about the location of the fault, so traces repair some failures that the baseline does not.
\end{hypothesis}

\begin{hypothesis}[Stalling]
\label{h: stall}
When the feedback does not localise the fault, the model often resubmits identical or near-identical code; traces should reduce this.
\end{hypothesis}
 
\begin{hypothesis}[Information floor]
\label{h: floor}
Where a failing test is hidden, the feedback carries at most one bit; no trace can be produced, so repair attempts should be no more productive than fresh samples.
\end{hypothesis}
 
\begin{hypothesis}[Ceiling]
\label{h: ceiling}
Traces help with implementation slips, such as a wrong condition, but not with a misread task, wrong algorithm or missing functionality.
Here the trace faithfully records a program that does the wrong thing.
\end{hypothesis}
 
\begin{hypothesis}[Usable information]
\label{h: usable}
Multiple traces added at once, in a single context window, are not necessarily more useful.
One trace should perform at least as well as all traces, at lower cost.
\end{hypothesis}
\begin{table*}[t]
\centering
\caption{Accuracy on all tasks (\%) and tasks repaired on the \bpd{} set (count). 
``Fresh'' makes six plain calls per task.
By construction, the baseline repairs none of the \bpd{} set.
Best performance in each row is shown in \textbf{bold}.}
\label{tab: main}
\small
\setlength{\tabcolsep}{5pt}
\begin{tabular}{lcc c cc cc cc}
\toprule
 & & & Baseline & \multicolumn{2}{c}{Fresh start} & \multicolumn{2}{c}{One trace} & \multicolumn{2}{c}{All traces} \\
\cmidrule(lr){5-6}\cmidrule(lr){7-8}\cmidrule(lr){9-10}
Benchmark & Model & \bpd{} set & Acc. (\%) & Acc. (\%) & Repaired & Acc. (\%) & Repaired & Acc. (\%) & Repaired \\
\midrule
\multirow{5}{*}{BigCodeBench} 
 & 7B & 320 & 63.5 & 70.9 & 84 & \textbf{71.7} & \textbf{93} & 70.2 & 76 \\
 & 14B & 258 & 71.3 & 76.9 & 64 & \textbf{78.8} & \textbf{85} & 77.3 & 68 \\
 & 32B & 222 & 76.3 & 80.3 & 45 & 81.8 & 63 & \textbf{82.1} & \textbf{66} \\
 & 72B & 214 & 76.3 & 79.3 & 34 & \textbf{81.4} & \textbf{58} & 80.8 & 51 \\
 & Pooled & 1,014 & 71.9 & 76.8 & 227 (22.4\%) & \textbf{78.4} & \textbf{299 (29.5\%)} & 77.6 & 261 (25.8\%) \\
\midrule
\multirow{5}{*}{HumanEval} 
 & 7B & 17 & 89.6 & \textbf{95.1} & \textbf{9} & \textbf{95.1} & \textbf{9} & 93.9 & 7 \\
 & 14B & 11 & 93.3 & 97.0 & 6 & 97.0 & 6 & \textbf{98.8} & \textbf{9} \\
 & 32B & 6 & 96.3 & 97.6 & 2 & \textbf{98.2} & \textbf{3} & 97.6 & 2 \\
 & 72B & 6 & 96.3 & \textbf{98.2} & \textbf{3} & \textbf{98.2} & \textbf{3} & 96.3 & 0 \\
 & Pooled & 40 & 93.9 & 97.0 & 20 (50.0\%) & \textbf{97.1} & \textbf{21 (52.5\%)} & 96.6 & 18 (45\%)\\
\midrule
\multirow{5}{*}{LiveCodeBench}
 & 7B & 121 & 27.4 & \textbf{32.0} & \textbf{8} & 29.1 & 3 & 29.7 & 4 \\
 & 14B & 120 & 28.6 & \textbf{33.7} & \textbf{9} & 32.6 & 7 & 30.9 & 4 \\
 & 32B & 109 & 33.7 & 38.3 & 8 & 38.3 & 8 & \textbf{38.9} & \textbf{9} \\
 & 72B & 99 & 34.9 & \textbf{37.1} & \textbf{4} & \textbf{37.1} & \textbf{4} & 36.6 & 3 \\
 & Pooled & 449 & 31.1 & \textbf{35.3} & \textbf{29 (6.5\%)} & 34.3 & 22 (4.9\%) & 34.0 & 20 (4.5\%) \\
\bottomrule
\end{tabular}
\end{table*}

\section{Methodology}
\label{sec: method}

\bpd{} extends a PyCapsule-style repair loop~\cite{adnan2025pycapsule} with one change.
When a program fails an assertion on a test whose input is visible, the feedback contains the program annotated with its own execution trace, showing the model what a developer would see with a breakpoint on every executed line.
All other failures receive the baseline's structured error message.

We run the failing test under a line-level tracer \texttt{sys.settrace} and write three kinds of annotations as end-of-line comments (Fig.~\ref{fig: trace}).
Assignments show the values taken by the assigned variable, loop headers show the sequence of loop-variable values, and control-flow statements show the iteration at which they were reached and which iterations were skipped as a result.
Sequences over 6 elements keep the first 4 and last 2 with $\dots$ between; values over 40 characters are clipped; and a trace is capped at 6,000 characters, with no trace if tracing runs past 20,000.
The prompt also contains the failing input, and the expected and actual outputs.
\emph{One trace} annotates the program with the first failing test only; \emph{all traces} adds a trace for every failing visible test sequentially.
 
\begin{figure}[h]
\begin{lstlisting}[style=trace]
def sum_of_evens(nums): # nums = [2, 4, 3, 2]
    total = 0           # total = 0
    for n in nums:      # n = 2, 4, 3
        if n % 2 == 1:  # Fx2, T
            break       # iter 3
        total += n      # total = 2, 6
    return total        # returns 6
\end{lstlisting}
\vspace{-0.6em}
{\footnotesize\ttfamily\ \ \ Input: [2, 4, 3, 2]\quad Expected: 8\quad Actual: 6}
\caption{An illustrative \bpd{} prompt fragment.}
\label{fig: trace}
\end{figure}
 
\bpd{} differs from LDB~\cite{zhong2024ldb} in three ways.
It is applied only to assertion failures, leaving crashes to tracebacks that already localise them.
It does not segment the program into blocks or ask for per-block verdicts, since function-level programs fit whole in the prompt.
And it places each value on the line that produced it rather than in a separate listing.

\section{Results}
\label{sec: results}


\subsection{Experimental Setup}
\label{subsec: result_setup}

\textbf{Benchmarks.}
We use HumanEval~\cite{codex} (164 tasks), BigCodeBench~\cite{zhuo2025bigcodebench} (1,140 tasks, full split) and 175 LiveCodeBench~\cite{jain2025livecodebench} problems (version~\cite{lcb_link}).
On HumanEval and BigCodeBench all tests are visible, and a failed assertion reports the input, the expected output and the actual output.
LiveCodeBench also has hidden tests, whose failure is reported only as ``a hidden test failed'', with no input or trace.
 
\textbf{Models.}
We use Qwen2.5 Instruct models with 7B, 14B, 32B and 72B parameters~\cite{qwen2025qwen25}, all released in December 2024, before the release of the LiveCodeBench benchmark version used in our study~\cite{lcb_link}.

\textbf{Protocol.}
The baseline is a PyCapsule-style loop~\cite{adnan2025pycapsule}: an initial generation (attempt~0) followed by up to five repair attempts with structured error messages, stopping when all tests pass.
The \bpd set comprises tasks for which the baseline exhausted all repair attempts without success, ending with an assertion failure.
For each, we rerun the full loop with \bpd{} feedback.
Total accuracy with \bpd{} is the baseline's passes plus the \bpd{} run's passes on the \bpd{} set.
Since rerunning selected failures may solve some problems by chance on the first attempt, we include a \emph{fresh-start} control that makes six independent plain calls per \bpd{}-set task, without feedback or iteration, following the fresh-start protocol of Debugging Decay Index~\cite{ddi}.
 
\textbf{Measures.}
We report accuracy as the share of a benchmark's tasks that are solved, the share of the \bpd{} set repaired (separating attempt~0), a public-tests-only view for LiveCodeBench, repair behaviour (identical resubmission, mention of ``trace'') and token usage with the Qwen2.5 tokeniser.

\subsection{RQ1: Which failures remain?}
\label{subsec: res_rq1}

Assertion failures account for 79.0\% of the failures left by the baseline on BigCodeBench, all 40 on HumanEval and 93.2\% on LiveCodeBench, with little variation across model sizes (Table~\ref{tab:residual}).
They also dominate the repair process itself.
Across all attempts of earlier PyCapsule runs on BigCodeBench, assertion errors make up 47--55\% of all error events, three to five times the next type, \texttt{ValueError}, and their share rises with capability, from 46.7\% for Qwen2.5-Coder 7B to 55.0\% for 32B.
As models improve, crashes become rarer and wrong answers make up more of what remains.
 
\begin{table}[h]
\centering
\caption{Share of residual failures that are assertion failures (\%) 
}
\label{tab:residual}
\small
\begin{tabular}{lrrrrr}
\toprule
Benchmark & 7B & 14B & 32B & 72B \\
\midrule
BigCodeBench  & 76.9 & 78.9 & 82.2 & 79.3 \\
HumanEval     & 100 & 100 & 100 & 100 \\
LiveCodeBench & 95.3 & 96.0 & 94.0 & 86.8 \\
\bottomrule
\end{tabular}
\end{table}

\subsection{RQ2: Do traces beat fresh sampling?}
\label{subsec: result_rq2}
 
Table~\ref{tab: main} reports the main results.
Against the baseline, one trace raises pooled accuracy from 71.9\% to 78.4\% on BigCodeBench, from 93.9\% to 97.1\% on HumanEval and from 31.1\% to 34.3\% on LiveCodeBench.
Much of this gain, however, is available without any feedback.
Six fresh samples alone repair 22.4\% of the BigCodeBench \bpd{} set, 50.0\% of HumanEval and 6.5\% of LiveCodeBench, tasks that the baseline loop had failed after six calls.
Repair loops that keep refining an unlucky first program forgo these solutions, in line with the exploration--exploitation view of repair~\cite{tang2024rex}.
Attempt~0 of a \bpd{} run is itself such a sample: it passes on 9.5\% of BigCodeBench tasks, against 9.4\% for a fresh call.
 
At an equal budget, traces help only where the failing input is visible and the tasks are not algorithmic.
On BigCodeBench, one trace repairs 299 tasks against 227 for fresh sampling, 
and it is ahead for every model.
On HumanEval, fresh sampling and one trace achieve comparable repair rates, and the set is too small to separate them.
On LiveCodeBench, fresh sampling is ahead of both arms (29 against 22 and 20), though not significantly ($p = 0.31$).
\ref{h: local} is therefore supported on BigCodeBench, where it is testable, and not on LiveCodeBench, as~\ref{h: floor} predicts.
 
The advantage of the trace on BigCodeBench grows with model size compared to fresh sampling.
Fresh sampling repairs fewer tasks as models grow, from 26.2\% of the set at 7B to 15.9\% at 72B, while one trace stays between 27\% and 32\%.
Two factors likely contribute.
Larger models fail fewer tasks, so the tasks they leave are harder, and their fresh samples are less productive: a single fresh call passes on 10.9\% of the 7B set but only 6.9\% of the 72B set.
A capable model sampled again without feedback tends to reproduce the same wrong program, so resampling adds little.
Feedback gives it something new to act on, and \bpd{}'s repair rate falls much less as the remaining tasks grow harder.

\subsection{RQ3: How do traces change repair behaviour?}
\label{subsec: result_rq3}

The clearest effect of the trace is that the model stops resubmitting the same code (Table~\ref{tab: behaviour}).
Identical resubmissions fall from 28\% to 8\% on BigCodeBench and from 23\% to 6\% on HumanEval, and similarity to the previous program falls from 0.96 to 0.91 and from 0.90 to 0.79, so the model makes larger edits.
These edits are more often productive: attempts after which more tests pass rise from 13\% to 17\% and from 12\% to 24\%, while attempts after which fewer pass stay at 7\% on BigCodeBench and rise from 11\% to 14\% on HumanEval.
A quarter to a third of one-trace replies refer to the trace by name.
This supports \ref{h: stall}.

\begin{table}[h]
\centering
\caption{Repair attempts on the \bpd{} set (BigCodeBench / HumanEval).}
\label{tab: behaviour}
\small
\setlength{\tabcolsep}{3pt}
\begin{tabular}{lccc}
\toprule
Measure & Baseline & One trace & All traces \\
\midrule
Identical resubmission (\%) & 28--23 & 8--6 & 9--5 \\
Similarity to previous & .96--.90 & .91--.79 & .90--.80 \\
More tests pass (\%) & 13--12 & 17--24 & 16--16 \\
Mentions ``trace'' (\%) & 0 & 25--32 & 19--15 \\
\bottomrule
\end{tabular}
\end{table}
 

A change in behaviour, however, does not guarantee a repair, and LiveCodeBench shows where it falls short.
LiveCodeBench differs from the other benchmarks in two ways: it has hidden tests, and its problems are algorithmic.
To separate the two, Table~\ref{tab: public} judges success only on the public tests, where the model sees the input, the expected output, and the trace.
Under iterative repair, 27.8\% of the \bpd{} set already passes the public tests and fails only hidden ones.
For these problems, each attempt returns a single bit, and no trace can be produced, which is the regime of~\ref{h: floor}.
Consistent with~\ref{h: floor}, repair attempts are no more productive than fresh samples here: fresh samples solve 29 tasks, more than either \bpd{} arm, and the trace-guided attempts after attempt~0 solve 11, the same number as attempt~0 alone.

\begin{table}[h]
\centering
\caption{LiveCodeBench \bpd{} set judged on public tests only: share cleared (\%), with the share cleared at attempt 0 in brackets.}
\label{tab: public}
\small
\begin{tabular}{lrrrr}
\toprule
Model & Set & Baseline & One trace & All traces \\
\midrule
7B  & 121 & 12.4 & 18.2 (12.4) & 19.8 (10.7) \\
14B & 120 & 30.0 & 29.2 (18.3) & 23.3 (14.2) \\
32B & 109 & 33.0 & 38.5 (15.6) & 33.0 (15.6) \\
72B & 99  & 38.4 & 29.3 (14.1) & 31.3 (19.2) \\
\midrule
Pooled & 449 & 27.8 & 28.5 (15.1) & 26.5 (14.7) \\
\bottomrule
\end{tabular}
\end{table}

The public tests show that hidden tests are not the whole explanation.
Where the trace is available, \bpd{} still gives no gain: the \bpd{} set clears its public tests in 27.8\% of problems under iterative repair, 28.5\% with one trace and 26.5\% with all traces, accuracy on all 175 problems is unchanged (51.7\%, 52.1\%, 50.9\%), and the 14B and 72B models lose.
A fresh first sample alone accounts for about half of the public tests cleared.
This supports~\ref{h: ceiling}.
A wrong answer in competitive programming more often reflects a wrong algorithm or a misread constraint than an implementation slip, and on such benchmarks LLMs are limited mainly by their understanding of the problem rather than by their ability to repair~\cite{zheng2025multiturn}.
The trace then faithfully records a program that does what it was written to do, so the larger edits it prompts are edits to the wrong idea.
Continuing to revise a program that already passes its public tests, in response to a one-bit hidden-test signal, may further explain the losses.

\subsection{RQ4: Is one trace enough, and what does it cost?}
\label{subsec: result_rq5}

\begin{table}[h]
\centering
\caption{Token usage on the \bpd{} set, pooled over models (Qwen2.5 tokeniser).
BPD is BigCodeBench, HE is HumanEval and LCB is LiveCodeBanch}
\label{tab: tokens}
\small
\setlength{\tabcolsep}{4pt}
\begin{tabular}{llrrrr}
\toprule
Bench. & Run & Calls & Tok./call & Total (k) & Change \\
\midrule
\multirow{3}{*}{BCB} & Baseline & 6,086 & 1,795 & 10,925 & \\
 & One trace & 5,095 & 2,061 & 10,502 & $-3.9\%$ \\
 & All traces & 5,203 & 2,692 & 14,007 & $+28.2\%$ \\
\midrule
\multirow{3}{*}{HE} & Baseline & 240 & 943 & 226 & \\
 & One trace & 167 & 1,139 & 190 & $-16.0\%$ \\
 & All traces & 168 & 1,604 & 270 & $+19.1\%$ \\
\midrule
\multirow{3}{*}{LCB} & Baseline & 2,694 & 1,511 & 4,070 & \\
 & One trace & 2,603 & 1,825 & 4,749 & $+16.7\%$ \\
 & All traces & 2,628 & 2,306 & 6,060 & $+48.9\%$ \\
\bottomrule
\end{tabular}
\end{table}

One trace repairs more tasks in total than all traces on every benchmark.
Fewer replies mention the trace when all traces are shown (Table~\ref{tab: behaviour}), suggesting the model attends less to longer traces.
This supports~\ref{h: usable} and agrees with the finding that the benefit of traces falls as they grow more complex~\cite{haque2025traces}.
 
A trace lengthens every call (Table~\ref{tab: tokens}), so the total cost depends on whether it saves calls by solving tasks sooner.
Total tokens factorise into calls times tokens per call, so a format that makes each call's token usage  $\rho$ times longer pays for itself only if it cuts calls by at least $1 - 1/\rho$.
For one trace, $\rho$ is 1.15--1.21, giving a break-even of 13--17\%.
One trace cut calls by 16\% on BigCodeBench and 30\% on HumanEval, lowering total tokens by 3.9\% and 16.0\%, but by only 3\% on LiveCodeBench, raising them by 16.7\%.
All traces, with $\rho$ of 1.50--1.70, never pay for themselves.
Fresh sampling is cheaper still per call, since its prompt carries no history. 

\section{Discussion, Limitations and Future Work}
\label{sec: discussion}
 
The results separate two conditions for a trace to help.
The first, that the feedback exposes a failing input, is necessary: under hidden-test verdicts no trace exists, and fresh sampling does at least as well (\ref{h: floor}).
On LiveCodeBench's public tests, the trace is visible and still gives no gain.
Our explanation is~\ref{h: ceiling}: competitive-programming failures are more often wrong algorithms or misread constraints than implementation slips. 
We support it only indirectly, and labelling failure causes is the natural next step.
The gain is modest: over the whole of BigCodeBench, one trace reaches 78.4\% against 76.8\% for fresh sampling, and much of what a repair loop recovers from its residual failures can be recovered by sampling again.
Traces matter more for larger models, whose residual failures resist resampling.
We offer this as a hypothesis, since we did not run LDB.
 
\textbf{Practical rules.}
(i) Trace assertion failures only.
(ii) Under hidden-test verdicts, and on competitive-programming tasks, spend the budget on fresh samples.
(iii) Show one trace, not all.
(iv) Check the break-even rule: a format that lengthens calls by $\rho$ must cut calls by $1 - 1/\rho$.
 
\textbf{Threats to validity.}
The \bpd{} set is selected for failure under iterative repair, so any rerun repairs some tasks by chance; the fresh-start control at an equal budget addresses this.
Each configuration was run once, so we cannot report run-to-run variance.
We classify failures by the exception raised, and resubmission, similarity and the word ``trace'' are coarse proxies for behaviour.
We study one model family and Python function-level tasks; the HumanEval \bpd{} set has only 40 tasks.
Longer programs may need block-level summaries of the kind used by LDB.

\section{Conclusion}
\label{sec:conclusion}
 
This study asked whether execution traces help an LLM repair the assertion failures that iterative repair leaves behind, and whether they do so better than sampling new programs at the same budget.
Assertion failures make up 79--100\% of the residual failures on three benchmarks, and our information view explains why they persist.

Our verdicts on the five hypotheses are as follows.
Traces localise faults that the error message does not (\ref{h: local}).
On BigCodeBench, where one trace repaired 26.7\% of the residual failures against 22.4\% for fresh sampling; on HumanEval the two were within two of 40 tasks, and on LiveCodeBench fresh sampling repaired more (6.5\% against 4.9\%).
Traces clearly change repair behaviour (\ref{h: stall}): identical resubmissions fell from 28\% to 8\% on BigCodeBench and from 23\% to 6\% on HumanEval, and more attempts increased the number of tests passed.
Where the failing test was hidden, repair attempts were no more productive than fresh samples (\ref{h: floor}).
Where the failing public test was visible on LiveCodeBench, the trace still gave no gain (28.5\% against 27.8\%), consistent with \ref{h: ceiling}.
One trace repaired at least as many tasks as all traces on every benchmark and was the only variant to lower token cost, by 3.9\% on BigCodeBench and 16.0\% on HumanEval (\ref{h: usable}).

The central lesson is that changing the model's behaviour is not the same as solving the problem.
A trace makes the model edit instead of stall, and on BigCodeBench those edits repair failures that resampling does not.
Competitive-programming problems, however, are more often failed because the model misunderstands the problem than because it implements it incorrectly~\cite{zheng2025multiturn}, and there, a trace only records a program built on the wrong idea.
Repair feedback should therefore be compared with fresh sampling at an equal budget, and a repair loop should resample rather than trace when the failing tests are hidden or the task is algorithmic.
Future work should label the causes of failure, extend the comparison to other model families and longer programs, and build loops that choose between tracing and resampling according to the feedback available.

\begingroup
\renewcommand{\thefootnote}{}
\footnotetext{{During the preparation of this manuscript, the authors used Claude Sonnet 5.5 to debug and refine the draft text. 
The authors have reviewed and edited all AI-assisted output and take full responsibility for the content of this publication.
The code and data are openly available at \url{https://github.com/TheOpenSI/SYNKRASIS/tree/BPDebug}}}
\endgroup



\bibliographystyle{IEEEtran}
\bibliography{references}

\begin{thebibliography}{10}
\providecommand{\url}[1]{#1}
\csname url@samestyle\endcsname
\providecommand{\newblock}{\relax}
\providecommand{\bibinfo}[2]{#2}
\providecommand{\BIBentrySTDinterwordspacing}{\spaceskip=0pt\relax}
\providecommand{\BIBentryALTinterwordstretchfactor}{4}
\providecommand{\BIBentryALTinterwordspacing}{\spaceskip=\fontdimen2\font plus
\BIBentryALTinterwordstretchfactor\fontdimen3\font minus \fontdimen4\font\relax}
\providecommand{\BIBforeignlanguage}[2]{{%
\expandafter\ifx\csname l@#1\endcsname\relax
\typeout{** WARNING: IEEEtran.bst: No hyphenation pattern has been}%
\typeout{** loaded for the language `#1'. Using the pattern for}%
\typeout{** the default language instead.}%
\else
\language=\csname l@#1\endcsname
\fi
#2}}
\providecommand{\BIBdecl}{\relax}
\BIBdecl

\bibitem{chen2021codex}
M.~Chen, J.~Tworek, H.~Jun, Q.~Yuan \emph{et~al.}, ``Evaluating large language models trained on code,'' \emph{arXiv preprint arXiv:2107.03374}, 2021.

\bibitem{zhuo2025bigcodebench}
T.~Y. Zhuo, M.~C. Vu, J.~Chim, H.~Hu, W.~Yu \emph{et~al.}, ``{BigCodeBench}: Benchmarking code generation with diverse function calls and complex instructions,'' in \emph{International Conference on Learning Representations}, 2025.

\bibitem{jain2025livecodebench}
N.~Jain, K.~Han, A.~Gu, W.-D. Li, F.~Yan, T.~Zhang, S.~Wang, A.~Solar-Lezama, K.~Sen, and I.~Stoica, ``{LiveCodeBench}: Holistic and contamination free evaluation of large language models for code,'' in \emph{International Conference on Learning Representations}, 2025.

\bibitem{chen2024selfdebug}
X.~Chen, M.~Lin, N.~Sch{\"a}rli, and D.~Zhou, ``Teaching large language models to self-debug,'' in \emph{International Conference on Learning Representations}, 2024.

\bibitem{adnan2025pycapsule}
M.~Adnan, Z.~Xu, and C.~C.~N. Kuhn, ``Large language model guided self-debugging code generation,'' \emph{arXiv preprint arXiv:2502.02928}, 2025.

\bibitem{zhong2024ldb}
L.~Zhong, Z.~Wang, and J.~Shang, ``Debug like a human: A large language model debugger via verifying runtime execution step by step,'' in \emph{Findings of the Association for Computational Linguistics: ACL 2024}, 2024, arXiv:2402.16906.

\bibitem{olausson2024selfrepair}
T.~X. Olausson, J.~P. Inala, C.~Wang, J.~Gao, and A.~Solar-Lezama, ``Is self-repair a silver bullet for code generation?'' in \emph{International Conference on Learning Representations}, 2024.

\bibitem{mapcoder}
\BIBentryALTinterwordspacing
M.~A. Islam, M.~E. Ali, and M.~R. Parvez, ``{M}ap{C}oder: Multi-agent code generation for competitive problem solving,'' in \emph{Proceedings of the 62nd Annual Meeting of the {A}ssociation for {C}omputational {L}inguistics (Volume 1: Long Papers)}.\hskip 1em plus 0.5em minus 0.4em\relax Bangkok, Thailand: Association for Computational Linguistics, Aug. 2024, pp. 4912--4944. [Online]. Available: \url{https://aclanthology.org/2024.acl-long.269/}
\BIBentrySTDinterwordspacing

\bibitem{mbpp}
J.~Austin, A.~Odena, M.~Nye, M.~Bosma, H.~Michalewski, D.~Dohan, E.~Jiang, C.~Cai, M.~Terry, Q.~Le \emph{et~al.}, ``Program synthesis with large language models,'' \emph{arXiv preprint arXiv:2108.07732}, 2021.

\bibitem{haque2025traces}
M.~Haque, P.~Babkin, F.~Farmahinifarahani, and M.~Veloso, ``Towards effectively leveraging execution traces for program repair with code {LLMs},'' in \emph{Proceedings of the 4th International Workshop on Knowledge-Augmented Methods for Natural Language Processing (KnowledgeNLP'25)}.\hskip 1em plus 0.5em minus 0.4em\relax Association for Computational Linguistics, 2025, pp. 160--179.

\bibitem{wang2025semantics}
\BIBentryALTinterwordspacing
J.~J. Wang, X.~Xie, Q.~Hu, S.~Liu, and Y.~Li, ``Do code semantics help? a comprehensive study on execution trace-based information for code large language models,'' in \emph{Findings of the {A}ssociation for {C}omputational {L}inguistics: {EMNLP} 2025}.\hskip 1em plus 0.5em minus 0.4em\relax Suzhou, China: Association for Computational Linguistics, Nov. 2025, pp. 10\,367--10\,385. [Online]. Available: \url{https://aclanthology.org/2025.findings-emnlp.548/}
\BIBentrySTDinterwordspacing

\bibitem{huang2026tracecoder}
J.~Huang, W.~Ye, W.~Sun, J.~Zhang, M.~Zhang, and Y.~Liu, ``{TraceCoder}: A trace-driven multi-agent framework for automated debugging of {LLM}-generated code,'' in \emph{Proceedings of the IEEE/ACM 48th International Conference on Software Engineering (ICSE)}, 2026, arXiv:2602.06875.

\bibitem{wu2026tracerepair}
J.~Wu, T.~Wu, M.~Zhang, Y.~Dong, and B.~Shen, ``Runtime execution traces guided automated program repair with multi-agent debate,'' \emph{arXiv preprint arXiv:2604.02647}, 2026.

\bibitem{voas1992pie}
J.~M. Voas, ``{PIE}: A dynamic failure-based technique,'' \emph{IEEE Transactions on Software Engineering}, vol.~18, no.~8, pp. 717--727, 1992.

\bibitem{ni2024next}
A.~Ni, M.~Allamanis, A.~Cohan, Y.~Deng, K.~Shi, C.~Sutton, and P.~Yin, ``{NExT}: Teaching large language models to reason about code execution,'' in \emph{Proceedings of the 41st International Conference on Machine Learning}, ser. PMLR, vol. 235, 2024, pp. 37\,929--37\,956.

\bibitem{huang2024cannot}
J.~Huang, X.~Chen, S.~Mishra, H.~S. Zheng, A.~W. Yu, X.~Song, and D.~Zhou, ``Large language models cannot self-correct reasoning yet,'' in \emph{International Conference on Learning Representations}, 2024.

\bibitem{dynafix2025}
Z.~Huang, L.~Xu, C.~Liu, W.~Sun, X.~Zhang, Y.~Lei, M.~Yan, and H.~Zhang, ``Dynafix: Iterative automated program repair driven by execution-level dynamic information,'' \emph{arXiv preprint arXiv:2512.24635}, 2025.

\bibitem{gu2024cruxeval}
A.~Gu, B.~Rozi{\`e}re, H.~Leather, A.~Solar-Lezama, G.~Synnaeve, and S.~I. Wang, ``{CRUXEval}: A benchmark for code reasoning, understanding and execution,'' in \emph{Proceedings of the 41st International Conference on Machine Learning}, 2024.

\bibitem{tang2024rex}
H.~Tang, K.~Hu, J.~P. Zhou, S.~C. Zhong, W.-L. Zheng, X.~Si, and K.~Ellis, ``Code repair with {LLMs} gives an exploration-exploitation tradeoff,'' in \emph{Advances in Neural Information Processing Systems}, 2024.

\bibitem{zheng2025multiturn}
K.~Zheng, J.~Decugis, J.~Gehring, T.~Cohen, B.~Negrevergne, and G.~Synnaeve, ``What makes large language models reason in (multi-turn) code generation?'' in \emph{International Conference on Learning Representations}, 2025.

\bibitem{cover2006elements}
T.~M. Cover and J.~A. Thomas, \emph{Elements of Information Theory}, 2nd~ed.\hskip 1em plus 0.5em minus 0.4em\relax Wiley, 2006.

\bibitem{xu2020usable}
Y.~Xu, S.~Zhao, J.~Song, R.~Stewart, and S.~Ermon, ``A theory of usable information under computational constraints,'' in \emph{International Conference on Learning Representations}, 2020.

\bibitem{shannon}
C.~E. Shannon, ``A mathematical theory of communication,'' \emph{The Bell System Technical Journal}, vol.~27, no.~3, pp. 379--423, 1948.

\bibitem{shi2023distracted}
F.~Shi, X.~Chen, K.~Misra, N.~Scales, D.~Dohan, E.~H. Chi, N.~Sch{\"a}rli, and D.~Zhou, ``Large language models can be easily distracted by irrelevant context,'' in \emph{Proceedings of the 40th International Conference on Machine Learning}, 2023.

\bibitem{liu2024lost}
N.~F. Liu, K.~Lin, J.~Hewitt, A.~Paranjape, M.~Bevilacqua, F.~Petroni, and P.~Liang, ``Lost in the middle: How language models use long contexts,'' \emph{Transactions of the Association for Computational Linguistics}, vol.~12, pp. 157--173, 2024.

\bibitem{ddi}
M.~Adnan and C.~C. Kuhn, ``Measuring and mitigating debugging effectiveness decay in code language models,'' \emph{Scientific Reports}, vol.~15, no.~1, p. 44120, 2025.

\bibitem{uesato2022process}
J.~Uesato, N.~Kushman, R.~Kumar, F.~Song, N.~Siegel, L.~Wang, A.~Creswell, G.~Irving, and I.~Higgins, ``Solving math word problems with process- and outcome-based feedback,'' \emph{arXiv preprint arXiv:2211.14275}, 2022.

\bibitem{lightman2024verify}
H.~Lightman, V.~Kosaraju, Y.~Burda, H.~Edwards, B.~Baker, T.~Lee, J.~Leike, J.~Schulman, I.~Sutskever, and K.~Cobbe, ``Let's verify step by step,'' in \emph{International Conference on Learning Representations}, 2024.

\bibitem{codex}
M.~Chen, J.~Tworek, H.~Jun, Q.~Yuan \emph{et~al.}, ``Evaluating large language models trained on code,'' \emph{arXiv preprint arXiv:2107.03374}, 2021.

\bibitem{lcb_link}
\BIBentryALTinterwordspacing
Test6, HuggingFace. [Online]. Available: \url{https://huggingface.co/datasets/livecodebench/code_generation_lite/blob/main/test6.jsonl}
\BIBentrySTDinterwordspacing

\bibitem{qwen2025qwen25}
{Qwen Team}, ``Qwen2.5 technical report,'' \emph{arXiv preprint arXiv:2412.15115}, 2024.

\end{thebibliography}


\end{document}